\documentclass[pra, floatfix, twocolumn, nofootinbib]{revtex4}
\usepackage{graphicx}
\usepackage{color}
\usepackage{amsmath, amsfonts, amssymb, bm}
\usepackage{gensymb}
\usepackage{braket}
\usepackage{subfigure}
\begin{document}
\title{Resonant single-photon double ionization driven by combined\\ intra- and interatomic electron correlations}
\author{A. Eckey}
\author{A. B. Voitkiv}
\author{C. M\"uller}
\affiliation{Institut f\"ur Theoretische Physik I, Heinrich Heine Universit\"at D\"usseldorf, Universit\"atsstr. 1, 40225 D\"usseldorf, Germany}
\date{\today}
\begin{abstract}
Double ionization of an atom by single-photon absorption in the presence of a neighbouring atom is studied. The latter is, first, resonantly photoexcited and, afterwards, transfers the excitation energy radiationlessly to the other atom, leading to its double ionization. The process relies on the combined effect of interatomic and intraatomic electron correlations. It can dominate over the direct double photoionization by several orders of magnitude at interatomic distances up to few nanometers. The relative position of the neighbouring atom is shown to exert a characteristic influence on the angular distribution of emitted electrons.
\end{abstract}
\maketitle

\section{Introduction}

Multiple photoionization of atoms and molecules belongs to the most sensitive probes of electron correlations \cite{SPDI-Review}. This is distinctly exemplified by the process of single-photon double ionization (SPDI) which would not exist in the absence of electron correlations. Since photoabsorption is induced by a one-body operator, the ejection of the second electron relies exclusively on the interaction among the electrons in the target.\footnote{We note that SPDI may also proceed through a shake-off mechanism \cite{SPDI-Review} which relies on the sudden termination of electron-electron interaction present in the initial bound state.} SPDI has most comprehensively been studied in helium, which represents the prototype atom for this process.

Apart from the direct channel of SPDI, double electron emission can also occur after resonant photoexcitation of an autoionizing state which lies in the double continuum \cite{resonant-SPDI}. This channel requires a more complex atomic structure than in helium. It has been studied thoroughly in alkaline-earth metal atoms (such as Be, Mg, Ca) which exhibit a heliumlike $ns^2$ electron configuration in the outer shell \cite{SPDI-alkali, SPDI-Mg-I, SPDI-Mg-II, SPDI-Ca-I}. Since the latter is well separated from the inner shells both spatially and energetically, these atoms may be considered as quasi-two-electron systems. Resonant SPDI has been measured, for example, around the $3p\to 3d$ resonance 
in Ca \cite{SPDI-Ca-exp} and the $2p\to 3d$ resonance in Mg \cite{SPDI-Mg-exp}.

Autoionization processes can also occur when an atom is not isolated in space but in close vicinity to another atom, so that the electronic structures at both centers are coupled by long-range interactions. 
For example, two-center electron correlations can drive radiationless decay of an inner-valence vacancy in one of the atoms, in case when a single-center Auger decay of this vacancy is energetically forbidden. In such a situation, the vacancy may decay by transferring excitation energy to the neighboring atom. This process of interatomic Coulombic decay (ICD) \cite{ICD, ICDres, ICDrev} can be much faster than a single-center radiative decay. It has been observed in a  variety of systems, comprising noble gas dimers \cite{dimers} and clusters \cite{clusters}. The closely related process of two-center photoionization (2CPI) was also studied, both theoretically \cite{2CPI, Perina} and experimentally in He-Ne dimers \cite{2CPIexp} and Ne-Ar clusters \cite{Hergenhahn}. A similar energy transfer process was identified in the photoionization of helium droplets doped with rare gas atoms \cite{droplets}.

Interatomic electron correlations may also lead to double ionization after single-photon absorption. Double ionization of helium dimers was observed after irratiation with $\approx 64$\,eV synchrotron photons \cite{SPDI-He2}. The process was shown to rely on a knock-off reaction: the first electron, which is photoejected from one of the helium atoms, kicks out the second electron from the other helium atom in an ($e$, $2e$) collision. Double ionization of magnesium in Mg-He clusters was demonstrated to be largely enhanced due to electron-transfer-mediated decay for photon energies above the helium ionization threshold \cite{SPDI-ETMD-theo, SPDI-ETMD-exp}. Also here, the photoabsorption first produces a He$^+$ ion which, subsequently, is neutralized via electron transfer from a Mg atom. Simultaneously, a second electron is ejected from Mg to keep the energy balance. Finally, double ICD has been predicted to occur in endohedral fullerenes \cite{DICD-theo}. For example, an excited Mg$^+$ ($2p^53s^2$) ion embedded in C$_{60}$ can decay into its ground state, forming a Mg$^+$($2p^63s)$@C$_{60}^{2+}$ complex. A related experiment has been conducted very recently on alkali dimers such as K-Rb attached to helium droplets \cite{DICD-exp}. Excitation energy from the helium 1$s$2$s$ state was transfered to the dimer, leading to double ionization with subsequent dissociation into K$^+$ and Rb$^+$.

\begin{figure}[t]  
\vspace{-0.25cm}
\begin{center}
\includegraphics[width=0.35\textwidth]{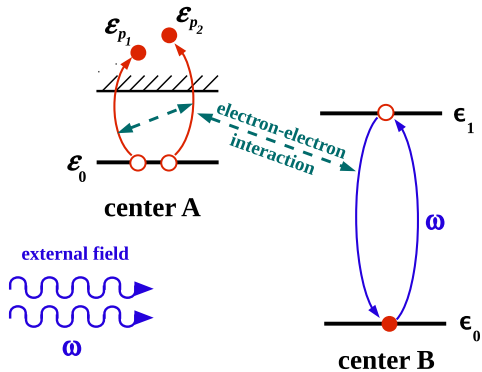}
\end{center}
\vspace{-0.5cm} 
\caption{Scheme of two-center single-photon double ionization. First, atom $B$ is resonantly photoexcited. Afterwards, upon radiationless energy transfer to atom $A$ via a two-center Auger decay, the latter is doubly ionized. The process involves three active electrons and relies on both interatomic and intraatomic electron correlations, as indicated by the dashed arrows.}
\label{figure1}
\end{figure}

In the present paper, we study another interatomic mechanism of SPDI, which may proceed in two-center heteroatomic systems. The process leads to double ionization of an atom $A$ in the presence of a neighboring atom $B$. It is schematically illustrated in Fig.~\ref{figure1}. The energy of the incident photon is assumed to be resonant with a dipole-allowed bound-bound transition in atom $B$ and to exceed the double ionization threshold of atom $A$. In a first step, atom $B$ is resonantly excited by photoabsorption, creating an autoionizing resonance state in the double continuum of the two-atomic system. Afterwards, the excitation energy is transfered via interatomic electron correlations to atom $A$ where it leads to double ionization due to intraatomic correlations between the two active electrons at center $A$.\footnote{In the context of ICD, the second step in two-center SPDI has been termed participator double resonant interatomic Coulombic decay (pDRICD); see Fig.~2\,a) in Ref.~\cite{ICDres}.} We show that the resonant two-center SPDI can dominate over the direct SPDI of atom $A$ by several orders of magnitude. It can even be substantially stronger than the direct single ionization of atom $A$. 

Our paper is organized as follows. In Sec.~II we present our theoretical considerations of two-center SPDI. After formulating the general framework, an analytical expression for the double-to-single ionization ratio will be obtained. In Sec.~IV we illustrate our findings by some numerical examples and discuss their physical implications.  Concluding remarks are given in Sec.~IV. Atomic units (a.u.) will be used throughout unless otherwise stated.

\section{Theory of two-center single-photon double ionization}

Let us consider a system consisting of two atoms, $A$ and $B$, separated 
by a sufficiently large distance $R$ such that their individuality 
is basically preserved. The atoms, which are initially in their ground states,  
are exposed to a resonant electromagnetic field. The latter 
will be treated as a classical electromagnetic wave of linear polarization, 
whose electric field component reads 
\begin{eqnarray}
{\bf F}({\bf r},t)= {\bf F}_0 \cos\left(\omega t - {\bf k} \cdot {\bf r}\right).
\label{field}
\end{eqnarray}
Here $\omega = c k $ and ${\bf k}$ 
are the angular frequency and wave vector,  
and ${\bf F}_0=F_0\,{\bf e}_z$ denotes the field strength vector
which is chosen to define the $z$ direction.

Assuming the atoms to be at rest,  
we take the position of the nucleus of atom $A$  
as the origin and denote the coordinates 
of the nucleus of atom $B$, the two (active) electrons of atom $A$ 
and that of atom $B$ by ${\bf R}$, ${\bf r}_j$ ($j\in\{1,2\}$) 
and ${\bf r}_3 ={\bf R} + \boldsymbol{\xi}$, 
respectively, where $\boldsymbol{\xi}$ 
is the position of the electron of atom $B$ 
with respect to its nucleus. 
Let atom $B$ have an excited state $\chi_e$
reachable from the ground state $\chi_g$ by a dipole-allowed transition. 

The total Hamiltonian describing the
two atoms in the external electromagnetic field reads 
\begin{eqnarray} 
H =  \hat{H}_0 + \hat{V}_{AB} + \hat{W},  
\label{hamiltonian}  
\end{eqnarray} 
where $ \hat{H}_0 $ is the sum of the Hamiltonians 
for the noninteracting atoms $A$ and $B$,  
$\hat{V}_{AB}$ the interaction between the atoms 
and $\hat{W} = \hat{W}_A + \hat{W}_B$ the interaction 
of the atoms with the electromagnetic field.
Within the dipole approximation and length gauge, the interaction $\hat{W}$ reads   
\begin{eqnarray} 
\hat{W} = \sum_{j=1,2,3} {\bf F}({\bf r}_j={\bf 0},t) \cdot {\bf r}_j\ .
\label{W} 
\end{eqnarray} 
The two terms with $j\in\{1,2\}$ compose the interaction $\hat{W}_A$ with the electrons in atom $A$,
whereas the term with $j=3$ the interaction $\hat{W}_B$ with the electron in atom $B$.
For electrons undergoing electric dipole transitions, the interatomic 
interaction reads 
\begin{eqnarray} 
\hat{V}_{AB} &=& \sum_{j=1,2}\left( \frac{{\bf r}_j\cdot \boldsymbol{\xi}}{R^3} 
- \frac{ 3 ({\bf r}_j\cdot{\bf R})(\boldsymbol{\xi}\cdot{\bf R})}{R^5} \right)\ .
\label{V_AB} 
\end{eqnarray} 
It is assumed that $\omega_{ge} R /c  \ll 1$, where $\omega_{ge}$ 
is the atomic transition frequency and $c$ the speed of light, such that 
retardation effects can be neglected.%
\footnote{Retardation effects in interatomic energy-transfer processes have 
recently been studied in Refs.~\cite{Buhmann,Salam}; see also Ref.~\cite{2CPI}.}

In the process of two-center SPDI one has essentially three different basic two-electron configurations, which are schematically illustrated in Fig.~\ref{figure1}:
(I) $\Psi_{g,g} = \Phi_{g}({\bf r}_1, {\bf r}_2) \chi_g(\boldsymbol{\xi})$ with total energy $E_{g,g} = \varepsilon_0 + \epsilon_1$, where both atoms are in the corresponding ground states $\Phi_g$ and $\chi_g$;
(II) $\Psi_{g,e} = \Phi_{g}({\bf r}_1, {\bf r}_2) \chi_e(\boldsymbol{\xi})$ with total energy $E_{g,e} = \varepsilon_0 + \epsilon_1$, in which atom $A$ is in the ground state while atom $B$ is in the excited state $\chi_e$;
(III) $\Psi_{{\bf p}_1,{\bf p}_2,g} = \Phi_{{\bf p}_1,{\bf p}_2}({\bf r}_1,{\bf r}_2) \chi_g(\boldsymbol{\xi})$ with total energy $E_{{\bf p}_1, {\bf p}_2, g} = \varepsilon_{p_1} + \varepsilon_{p_2} + \epsilon_0$, where both electrons of atom $A$ have been emitted into the continuum with asymptotic momenta ${\bf p}_1$ and ${\bf p}_2$, while the electron of atom $B$ has returned to the ground state. 

Within the second order of time-dependent perturbation theory, the probability amplitude for two-center SPDI can be written as
\begin{eqnarray}
S^{(2)}_{{\bf p}_1,{\bf p}_2}\! &=&\! -\int_{-\infty}^{\infty} dt\, \langle \Psi_{{\bf p}_1,{\bf p}_2,g}|\hat{V}_{AB}| \Psi_{g,e} \rangle\, e^{-i(E_{g,e}-E_{{\bf p}_1,{\bf p}_2,g})t}\nonumber\\
& & \times \int_{-\infty}^{t} dt'\, \langle \Psi_{g,e}|\hat{W}_{B}| \Psi_{g,g} \rangle\, e^{-i(E_{g,g}-E_{g,e})t'}
\label{S1}
\end{eqnarray}

By performing the inner time integral, we obtain
\begin{eqnarray} 
S^{(2)}_{{\bf p}_1,{\bf p}_2} &=& -i\int_{-\infty}^{\infty} dt\, \langle \Psi_{{\bf p}_1,{\bf p}_2,g}|\hat{V}_{AB}| \Psi_{g,e} \rangle\,\frac{F_0}{2} \nonumber\\
& & \times\,\frac{\langle \chi_e|\xi_z| \chi_g \rangle}{\epsilon_0+\omega-\epsilon_1+\frac{i}{2}\Gamma}\,e^{-i(E_{g,g}+\omega-E_{{\bf p}_1,{\bf p}_2,g})t}\ .\nonumber\\
\label{S2}
\end{eqnarray}
Here we have applied the rotating wave approximation and inserted the total width $\Gamma=\Gamma_r+\Gamma_a$ of the excited state $\chi_e$ in atom $B$. It accounts for the finite lifetime of this state and consists of the radiative width $\Gamma_r$ and the two-center Auger (ICD) width $\Gamma_a$. Taking also the outer time integral, we arrive at
\begin{eqnarray} 
S^{(2)}_{{\bf p}_1,{\bf p}_2} &=& -i\pi\, \langle \Phi_{{\bf p}_1,{\bf p}_2}|({\bf r}_1 + {\bf r}_2)| \Phi_g \rangle\cdot\left( {\bf e}_z - \frac{3R_z}{R^2}\,{\bf R}\right) \nonumber\\
& & \times\ \frac{F_0}{R^3}\, \frac{\left| \langle \chi_e|\xi_z| \chi_g \rangle \right|^2}{\Delta+\frac{i}{2}\Gamma}\ \delta(\varepsilon_{p_1} + \varepsilon_{p_2} - \varepsilon_0 - \omega)\ .\nonumber\\
\label{S3}
\end{eqnarray}
where the detuning from the resonance $\Delta = \epsilon_0+\omega-\epsilon_1$ has been introduced.
The delta function in Eq.~\eqref{S3} displays the law of energy conservation in the process.

From the transition amplitude we can obtain the fully differential ionization cross section in the usual way by taking the absolute square and dividing it by the interaction time $\tau$ and the incident flux $j=\frac{cF_0^2}{8\pi \omega}$, that is
\begin{eqnarray} 
\frac{{\rm d}^6\sigma^{(2)}}{{\rm d}^3p_1\,{\rm d}^3p_2} = \frac{1}{(2\pi)^6 j\tau}\,\left|S^{(2)}_{{\bf p}_1,{\bf p}_2}\right|^2\ .
\label{CS}
\end{eqnarray}
The factor $(2\pi)^{-6}$ arises from the fact that the continuum states in our calculations are normalized to a quantization volume of unity. Performing the integration over $\varepsilon_{p_2}$ with the help of the $\delta$-function in Eq.~\eqref{S3}, we obtain the five-fold differential cross section
\begin{eqnarray} 
\frac{{\rm d}^5\sigma^{(2)}}{{\rm d}\varepsilon_{p_1}{\rm d}^2\Omega_{p_1}{\rm d}^2\Omega_{p_2}} &=& \frac{\omega\,p_1\,p_2}{(2\pi)^4 c R^6}\ \frac{\left| \langle \chi_e|\xi_z| \chi_g \rangle \right|^4}{\Delta^2+\frac{1}{4}\Gamma^2}\nonumber\\
& & \hspace{-2cm} \times\, \big| \langle \Phi_{{\bf p}_1,{\bf p}_2}|({\bf r}_1 + {\bf r}_2)| \Phi_g \rangle \cdot\left( {\bf e}_z - 3\cos\theta_R\,{\bf e}_R\right) \big|^2\ ,\nonumber\\
\label{5CS}
\end{eqnarray}
where we have introduced the unit vector ${\bf e}_R={\bf R}/R$ along the internuclear separation and the angle $\theta_R$ between ${\bf R}$ and the field direction.

We can draw a comparison with the direct SPDI of atom $A$ by the electromagnetic field. The corresponding probability amplitude in the first order of perturbation theory is given by 
\begin{eqnarray}
S^{(1)}_{{\bf p}_1,{\bf p}_2} &=& -i\int_{-\infty}^{\infty} dt\, \langle \Phi_{{\bf p}_1,{\bf p}_2}|\hat{W}_A| \Phi_g \rangle\, e^{-i(\varepsilon_g-\varepsilon_{{\bf p}_1,{\bf p}_2})t}\nonumber\\
&=& -i\,\frac{F_0}{2}\, \langle \Phi_{{\bf p}_1,{\bf p}_2}|({\bf r}_1 + {\bf r}_2)\cdot{\bf e}_z| \Phi_g \rangle \nonumber\\
& & \times\ 2\pi\delta(\varepsilon_{p_1} + \varepsilon_{p_2} - \varepsilon_0 - \omega)\ .
\label{S-1C}
\end{eqnarray}
For the special cases, when the separation vector ${\bf R}$ between the atoms $A$ and $B$ is oriented either along the field direction or perpendicular to it, we can cast Eq.\,\eqref{S3} into the form
\begin{eqnarray} 
S^{(2)}_{{\bf p}_1,{\bf p}_2} &=& \frac{\alpha}{R^3}\,\frac{\left| \langle \chi_e|\xi_z| \chi_g \rangle \right|^2}{\Delta+\frac{i}{2}\Gamma}\,S^{(1)}_{{\bf p}_1,{\bf p}_2}
\label{S4}
\end{eqnarray}
where $\alpha=-2$ for ${\bf R}\parallel {\bf F}_0$ and $\alpha=1$ for ${\bf R}\perp {\bf F}_0$. When the field is exactly resonant with the transition $\chi_g\to\chi_e$ in atom $B$ and the interatomic distance is sufficiently large, so that $\Gamma_a\ll\Gamma_r$, this expression becomes 
\begin{eqnarray} 
S^{(2)}_{{\bf p}_1,{\bf p}_2} &=& \frac{3\alpha}{2i} \left(\frac{c}{\omega R}\right)^3\,S^{(1)}_{{\bf p}_1,{\bf p}_2}\ .
\label{S5}
\end{eqnarray}
Here we have used that the radiative width is given by $\Gamma_r=\frac{4\omega_{ge}^3}{3c^3}\left| \langle \chi_e|\xi_z| \chi_g \rangle \right|^2$ with $\omega_{ge}=\epsilon_1-\epsilon_0=\omega$. 

From Eq.\,\eqref{S5} we can infer that the cross section for two-center SPDI can be largely enhanced by a factor $[c/(\omega R)]^6\gg 1$ as compared with the usual one-center process of SPDI, where the neighboring atom $B$ is not involved. For example, assuming $\omega\approx 21.2$\,eV (corresponding to the first excitation energy in helium) and $R=10$\,\AA, an enormous enhancement by 6 orders of magnitude results. This implies further that two-center SPDI can even exceed the direct single ionization of atom $A$ considerably. Because the double-to-single ionization ratio is typically of the order $10^{-2}$, an enhancement by four orders remains. We point out that, nevertheless, two-center SPDI is not the dominant ionization channel because {\it single} ionization of atom $A$ via 2CPI is considerably stronger, being enhanced over direct single photoionization by a factor $[c/(\omega R)]^6$ as well \cite{2CPI}. The ratio of two-center SPDI--to--2CPI is thus of the order $10^{-2}$, just like the ratio between the corresponding one-center processes.

The processes of two-center SPDI and direct SPDI lead to the same final state, since atom $B$ eventually returns to its ground state and thus serves as a catalyzer. Therefore, the corresponding probability amplitudes  \eqref{S3} and \eqref{S-1C} are generally subject to quantum interference. However, for parameters where the two-center channel strongly dominates, the interference is of minor importance and may be neglected.

\section{Numerical Results and Discussion}

Next we illustrate some characteristic properties of two-center SPDI by numerical examples. Our focus lies on the angular distribution of the two electrons emitted from atom $A$, which strongly depends on the position ${\bf R}$ of the neighboring atom $B$, as we will show. 

\begin{figure*}[htb]
	\begin{center}
	\includegraphics[width=15cm]{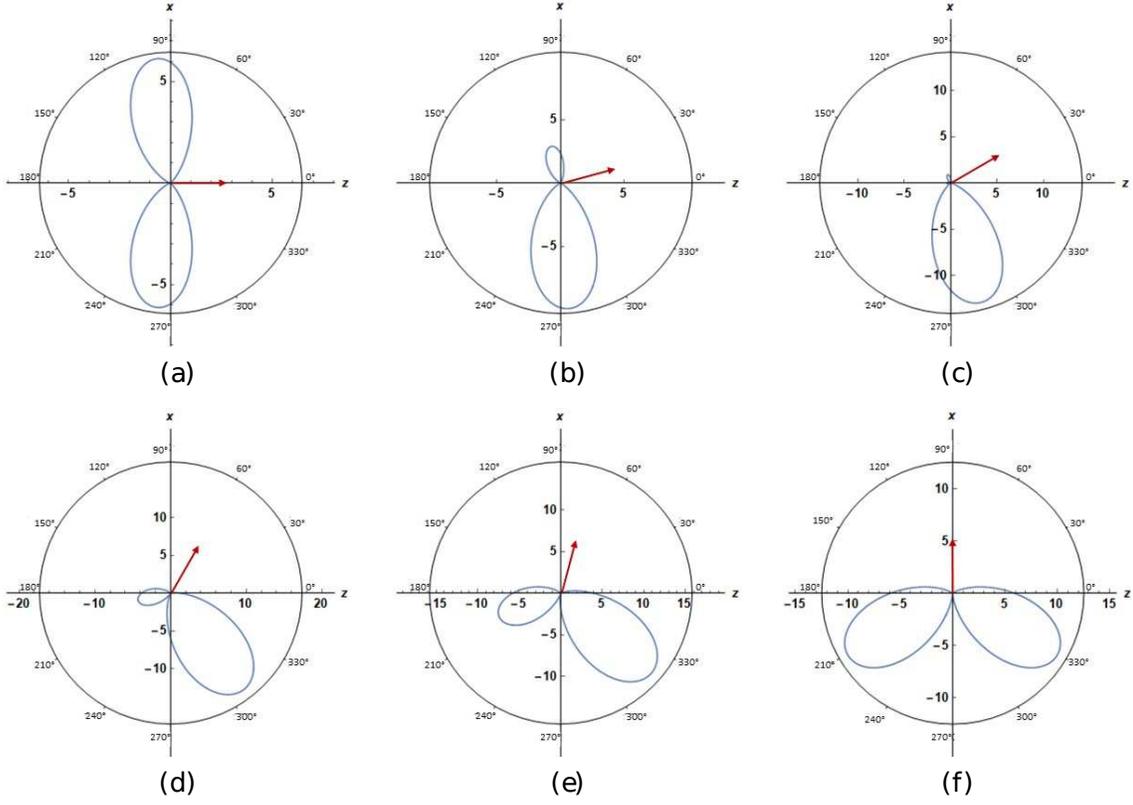}
		\caption{Five-fold differential cross section \eqref{5CS} of two-center SPDI in a He-Li$^{2+}$ system (in units of 10$^{-3}$\,a.u.). The photon energy, $\omega=91.8$\,eV, is resonant with the $1s\to2p$ transition in Li$^{2+}$ and the field strength is $F_0=10^{-5}$\,a.u. ($\approx 5\times 10^4$\,V/cm). The atoms are separated along the field direction by $R=10$\,a.u. From (a)\,-\,(c) in the upper row the polar emission angle $\theta_1$ of the first electron is varied from $0^{\circ}$ to $45^{\circ}$ in steps of $15^{\circ}$, as indicated by the red arrows. Similarly, from (d)\,-\,(f) in the lower row it is varied further from $60^{\circ}$ to $90^{\circ}$. The blue line shows a polar plot of the angular distribution of the second electron. Both electrons have the same energies $\varepsilon_{p_1}=\varepsilon_{p_2}\approx 6.8$\,eV and azimuthal angles $\varphi_1 = \varphi_2 = 0^{\circ}$.}
		\label{Fig2}
	\end{center}
\end{figure*}

A simple diatomic system is used as generic model to illustrate the effects. Since most of the studies on SPDI have considered helium atoms, we also use helium as atom $A$ in our model system. For the helium ground state, we apply a radially correlated wave function of the form
\begin{eqnarray}
\Phi_g({\bf r}_1,{\bf r}_2) = N\left( e^{-\alpha r_1}e^{-\beta r_2} + e^{-\alpha r_2}e^{-\beta r_1}\right)\ ,
\end{eqnarray}
with $\alpha=1.18853$, $\beta=2.18317$ determined from a variational calculation \cite{Keller} and $N$ denoting the normalization constant. The corresponding ground state energy amounts to $\varepsilon_0\approx-2.876$\,a.u. which differs from the exact value by 1\,\%. To describe the double continuum of helium we employ 2C wave functions, consisting of a symmetrized product of two Coulomb waves $\psi^{(-)}_{{\bf p}_j}({\bf r}_j)$ \cite{LandauQM} with nuclear charge $Z_A=2$. Final-state correlations are accounted for by inclusion of a Gamov factor $G({\bf p}_1-{\bf p}_2)$ \cite{LandauQM} which depends on the relative momentum between the electrons:
\begin{eqnarray}
\Psi_{{\bf p}_1,{\bf p}_2}({\bf r}_1,{\bf r}_2)\!\! &=&\!\! \frac{1}{\sqrt{2}}\Big[ \psi^{(-)}_{{\bf p}_1}({\bf r}_1)\psi^{(-)}_{{\bf p}_2}({\bf r}_2)+\psi^{(-)}_{{\bf p}_1}({\bf r}_2)\psi^{(-)}_{{\bf p}_2}({\bf r}_1) \Big]\nonumber\\
& & \times\ G({\bf p}_1-{\bf p}_2)\ .
\end{eqnarray}
Into the fully differential ionization cross section, the absolute square $|G({\bf p})|^2=\frac{2\pi}{p}\left( {\rm e}^{2\pi/p}-1\right)^{-1}$ of the Gamov factor enters.
We note that calculations of one-center double ionization of helium using these wave functions show good agreement with more sophisticated theories in terms of the shape of the photoelectron angular distributions (see, e.g., \cite{Keller, Maulbetsch1993, Macri}; the {\it absolute} value of the differential cross section is not well reproduced, though).
Application of these wave functions offers the advantage that the matrix elements in Eq.~\eqref{S3} can be evaluated analytically by standard means.

\begin{figure*}[htb]
	\begin{center}
		\includegraphics[scale=0.95]{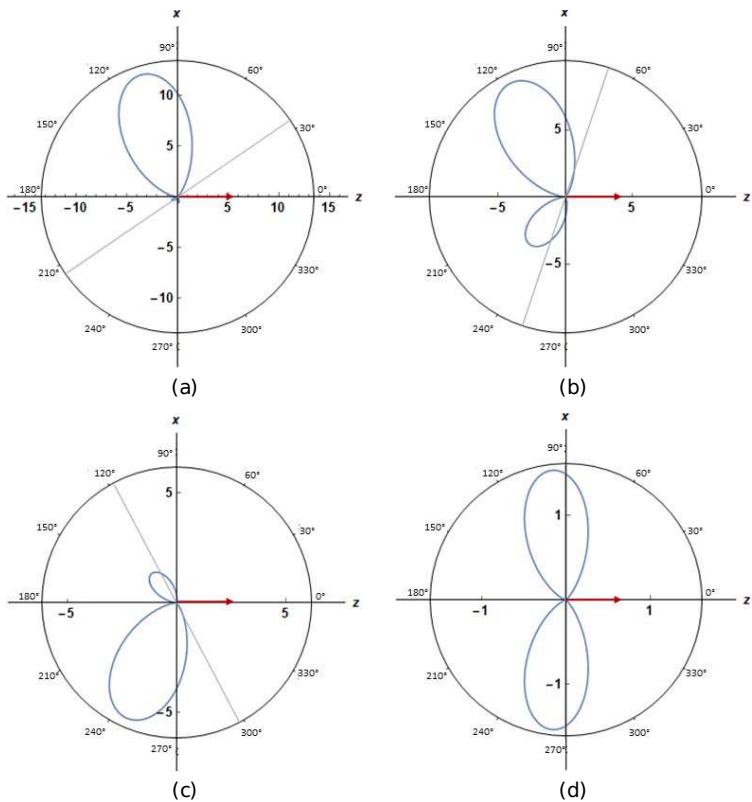}
		\caption{Same as Fig.~\ref{Fig2} but this time the polar emission angle of the first electron is kept fixed ($\theta_1=0^\circ$) while the orientation of the Li$^{2+}$ ion relative to the He atom varies: (a) $\theta_R=22.5^\circ$, (b) $\theta_R=45^\circ$, (c) $\theta_R=67.5^\circ$, (d) $\theta_R=90^\circ$. The straight grey lines show the resulting effective polarization axis [see Eq.~\eqref{eff-pol}]. }
		\label{Fig3}
	\end{center}
\end{figure*}

In order to provide sufficient energy to overcome the double ionization threshold of helium, a Li$^{2+}$ ion is assumed to constitute the neighboring atomic center $B$ in our generic model system. The latter is hydrogenlike with nuclear charge $Z_B=3$. The photon energy $\omega=\epsilon_1-\epsilon_0=3.375$\,a.u. is assumed to be resonant with the $1s\to2p$ transition in Li$^{2+}$. For the internuclear distance the value $R=10$\,a.u. is taken throughout. At this distance, the radiative width largely exceeds the two-center Auger width, so that $\Gamma\approx \Gamma_r$. The relative enhancement between two-center SPDI in He-Li$^{2+}$ versus ordinary one-center SPDI of helium amounts to $[c/(\omega R)]^6\sim 4.4\times 10^3$.
 
We have calculated photoelectron angular distributions in the case of coplanar emission, with the azimuthal angles of both ejected electrons set to $\varphi_1 = \varphi_2 = 0$. Their polar angles $\theta_1$ and $\theta_2$ are measured with respect to the polarization direction of the external field \eqref{field}, which was chosen to define the $z$ direction. We assume equal energy sharing between the electrons. 

Figure 2 shows polar plots of the five-fold differential cross section \eqref{5CS} when the interatomic separation vector lies parallel to the field axis (${\bf R}\parallel {\bf F}_0$). In panels (a)\,-\,(f) the polar emission angle of the first electron is varied stepwise from $0^{\circ}$ to $90^{\circ}$. We observe that the angular distributions closely agree with those known from one-center SPDI of helium, as a comparison with results from the literature reveals (see, e.g., Fig.~1 in \cite{Brauning} for $\theta_1\in\{0^{\circ}, 30^{\circ}, 60^{\circ}, 90^{\circ}\}$ and Fig.~4\,(c) in \cite{Thumm} for $\theta_1\approx \pm75^{\circ}$). With the rotation of the emission angle $\theta_1$ of the first electron, the angular distribution of the second electron changes accordingly. The repulsion of the two electrons becomes apparent this way, but it does not lead to back-to-back emission. Instead a preferred relative angle between $\theta_1$ and $\theta_2$ of about $100^\circ$--$140^\circ$ appears. 
For symmetry reasons, during ionization along one of the coordinate axes ($\theta_1=0^\circ$ or 90$^\circ$), two emission directions are equally probable for the second electron. For emission angles $\theta_1$ in between, one direction is preferred. 

The angular emission patterns are modified when the relative orientation between the atoms is changed. This is displayed in Fig.~\ref{Fig3} where the first electron is always ejected under $\theta_1=0^\circ$ and the internuclear angle $\theta_R$ is varied. A very strong sensitivity on the position of the Li ion is found.

\begin{figure*}[htbp]
	\begin{center}
		\includegraphics[scale=0.97]{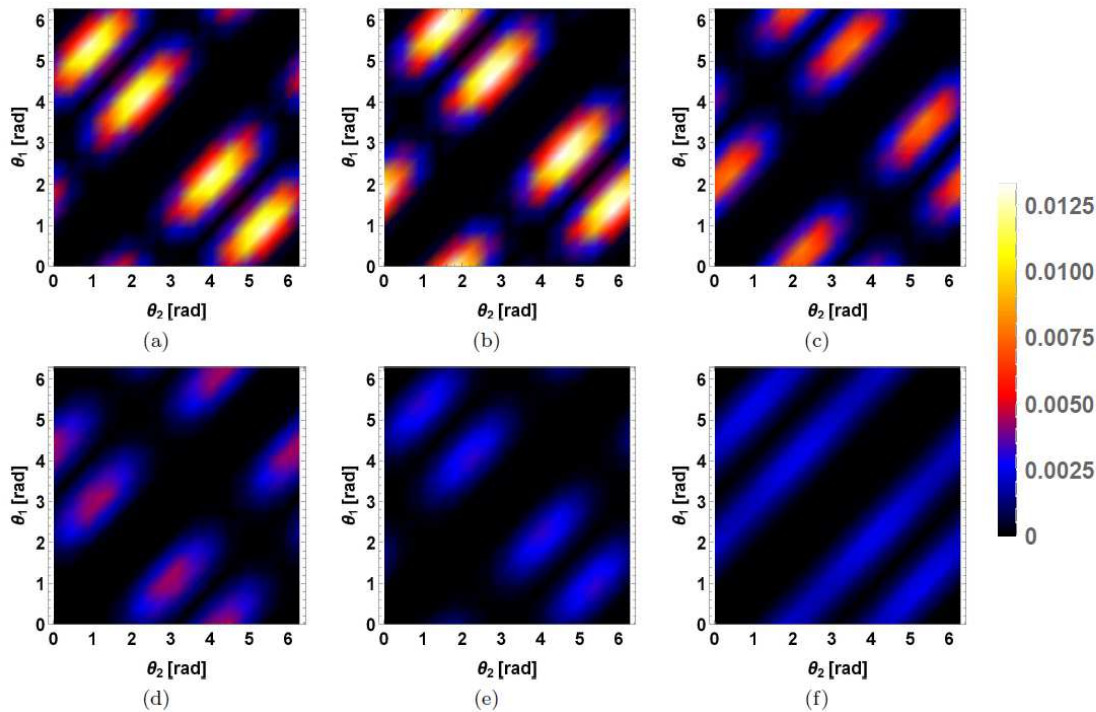}
		\caption{Density plots of the five-fold differential cross section \eqref{5CS} for two-center SPDI in a He-Li$^{2+}$ system, depending on the polar emission angles of both electrons. From (a)\,-\,(e) the relative interatomic orientation is varied from $\theta_R=0^\circ$ to $90^\circ$ in steps of $22.5^\circ$. Panel (f) shows the joint angular distribution when an average over the interatomic orientation is taken. The other parameters are as in Fig.~\ref{Fig2}.}
		\label{Fig4}
	\end{center}
\end{figure*}

The shapes of the photoelectron angular distributions can be understood by inspection of Eqs.~\eqref{S3} and \eqref{S-1C}. In case of one-center SPDI, the interaction operator $({\bf r}_1 + {\bf r}_2)\cdot{\bf e}_z$ appears, with the field polarization vector ${\bf e}_z$, which leads to the emission patterns known for this process. In case of two-center SPDI, a similar interaction operator arises, but the field polarization is replaced by an effective polarization vector
\begin{eqnarray}
	{\bf n}_{\text{eff}}={\bf e}_z - 3\cos\theta_R\,{\bf e}_R
	\label{eff-pol}
\end{eqnarray}
which represents a linear combination of the field polarization and the interatomic orientation vector.
Note that ${\bf n}_{\text{eff}}$ is generally not normalized but has length $|{\bf n}_{\text{eff}}|=(1+3\cos^2\theta_R)^{1/2}$. This effective polarization vector encodes the impact of the relative position of atom $B$ on the angular distribution of two-center SPDI. 

The angular spectra in Fig.~\ref{Fig2} refer to $\theta_R=0^\circ$, where ${\bf n}_{\text{eff}}=-2\,{\bf e}_z$. Thus, the effective polarization is (anti)parallel to the field polarization which explains, why the photoelectron emission patterns for two-center and one-center SPDI have identical shapes. 

Conversely, in Fig.~\ref{Fig3}\,(a)\,-\,(c) the effective polarization direction differs from the field polarization and, thus, from the emission direction of the first electron, as indicated by the straight grey lines. The angle between ${\bf e}_z$ and the effective polarization {\it axis} amounts to about 30$^\circ$ in panel (a), 75$^\circ$ in panel (b), and 60$^\circ$ in panel (c). Accordingly, Fig.~\ref{Fig3}\,(a) resembles Fig.~\ref{Fig2}\,(c), Fig.~\ref{Fig3}\,(b) resembles Fig.~\ref{Fig2}\,(e), and Fig.~\ref{Fig3}\,(c) resembles Fig.~\ref{Fig2}\,(d), provided a proper rotating reflection is applied. [In fact, if $\theta_R$ is changed and at the same time $\theta_1$ is laid along the corresponding effective polarization direction, an angular distribution equivalent to Fig.~2\,(a) results, which is only rotated by the angle $\theta_R$ (not shown).]
For $\theta_R=90^\circ$ [see Fig.~\ref{Fig3}\,d)], one has ${\bf n}_{\text{eff}}={\bf e}_z$ and the emission pattern coincides with the one in Fig.~\ref{Fig2}\,(a). Only the absolute values of the five-fold differential cross sections differ by a factor of 4, because the effective polarization vector in Fig.~\ref{Fig2}\,(a) has length $|{\bf n}_{\text{eff}}|=2$.

Our findings are confirmed by Fig.~\ref{Fig4} which shows two-dimensional density plots of the joint angular distribution of both electrons. In panels (a)\,-\,(e) the interatomic orientation is stepwise changed from $\theta_R=0^\circ$ to $90^\circ$. The emission pattern in Fig.~\ref{Fig4}\,(a), where ${\bf R}\parallel {\bf F}_0$ and thus ${\bf n}_{\text{eff}}=-2\,{\bf e}_z$, agrees with the corresponding result for one-center SPDI (see Fig.~4\,(a) in \cite{Thumm}). 
Four pronounced emission peaks appear which lie symmetrically to the vertical line defined by $\theta_1 + \theta_2 = 360^\circ$. These four peaks shift gradually as $\theta_R$ increases in panels (b)\,-(d) and eventually reach their initial positions in Fig.~\ref{Fig4}\,(e), where ${\bf R}\perp {\bf F}_0$ and thus ${\bf n}_{\text{eff}}={\bf e}_z$ holds. Consequently, the distributions in panels (a) and (e) are identical in shape and only differ in absolute value by a factor of 4 because of the different values of $|{\bf n}_{\text{eff}}|$.

In an experimental situation it can be difficult to fix the interatomic orientation between atoms $A$ and $B$. If, for example, a gas of van-der-Waals dimers $A$-$B$ was used, the internuclear axis would be randomly oriented. Therefore, we have averaged our results over the interatomic orientation, which results in the angular distribution in Fig.~\ref{Fig4}\,(f). The four emission peaks are smeared out into four parallel emission stripes which have the same slope and now extend over the complete plot range. 

Our calculations in this section were performed considering a He-Li$^{2+}$ model system in order to demonstrate some general features of two-center SPDI angular distributions. One should note, however, that it is very difficult to prepare such a system for conducting a dedicated experiment. Besides, the presence of the doubly ionized lithium ion will cause energy shifts in the neighbouring helium atom, which were not taken into account here.

In view of a potential experimental observation of two-center SPDI one could employ an earth-alkaline metal, such as calcium or magnesium, as atom $A$ in conjunction with helium as atom $B$. Using photon energies $\omega=21.2$\,eV (23.1\,eV), which are resonant to the $1s\to 2p$ ($1s\to 3p$) transition in helium, double ionization of Ca (Mg) could be probed. The corresponding thresholds are 18.0\,eV and 22.7\,eV, respectively \cite{NIST}. These target systems could be realized, for example, by Ca (or Mg) atoms attached to helium droplets (see, e.g., \cite{Ca-He} for a related experiment). Due to the smaller energy transfers and interatomic distances involved, the relative enhancement between two-center and one-center SPDI would be much larger here than in the He-Li$^{2+}$ model system considered above.

\section{Conclusion}
Single-photon double ionization in two-center systems of atoms $A$ and $B$ was studied, where resonant photoexcitation of $B$ leads to double ionization of $A$ via the combined action of interatomic and intraatomic electron-electron correlations. It was shown that, due to its resonant character, two-center SPDI can largely dominate over the ordinary one-center SPDI of atom $A$ for interatomic distances up to few nanometers. 

The influence of atom $B$ was also revealed in the angular distributions of emitted photoelectrons. They depend sensitively on the direction of an effective polarization vector, which arises as linear combination of the external field polarization and the interatomic separation vector. This effective polarization direction determines the shape of angular distributions from two-center SPDI in the same way as the field polarization vector does in case of one-center SPDI. The corresponding effects were illustrated by considering a simple two-center model system.

Our predictions could be tested, for example, in Ca-He or Mg-He systems, which possess low double-ionization thresholds. When the interatomic orientation is not fixed but randomly distributed in the target, the correspondingly averaged photoelectron angular distribution of two-center SPDI looks qualitatively different from the one-center case. This characteristic feature could help to experimentally identify the two-center process, in addition to its strong resonant enhancement over one-center SPDI in terms of the total cross section.

\section*{Acknowledgement}
This work has been funded by the Deutsche Forschungsgemeinschaft (DFG, German Research Foundation) under Grant No. 349581371 (MU 3149/4-1 and VO 1278/4-1). 


\end{document}